\documentclass[11pt,a4paper]{article}
\usepackage[body={16.5cm,25cm}]{geometry}
\usepackage[dvips]{color}
\usepackage{array}
\usepackage{stackrel}
\usepackage{tensor}
\usepackage{mathtools}
\usepackage[titletoc]{appendix}
\usepackage{fullpage}
\usepackage{amsmath, amsthm, amssymb}
\usepackage{centernot}
\usepackage{cite}
\usepackage[titletoc]{appendix}
\newcommand{\ds}{\displaystyle}
\renewcommand{\author}[1]{\large\rm #1\\ \bigskip}
\newcommand{\address}[1]{{\normalsize\it #1\\}\bigskip}
\renewcommand{\title}[1]{\bigskip\bigskip\Large\bf #1\bigskip\bigskip\\}

\newcommand{\beq}{\begin{equation}}
\newcommand{\eeq}{\end{equation}}
\newcommand{\Z}{{\mathbb Z}}
\newcommand{\C}{{\mathbb C}}

\newcommand{\slthree}{{\mathfrak sl_{3}}}

\newcommand{\phiss}{\tensor*[_{2}]{{\phi}}{_1}}
\newcommand{\phist}{\tensor*[_{4}]{{\phi}}{_3}}
\newcommand{\al}{{\alpha}}
\newcommand{\be}{{\beta}}
\newcommand{\ga}{{\gamma}}
\DeclareMathAlphabet{\mathbbmsl}{U}{bbm}{m}{sl}
\newcommand{\bs}[1]{{\boldsymbol{#1}}}

\newtheorem{proposition-definition}[theorem]{Proposition-Definition}
\usepackage{braket}

\theoremstyle{definition}

\theoremstyle{remark}

\numberwithin{equation}{section}

\begin{document}

\vglue 2 cm
\begin{center}
\title{Construction of  $R$-matrices
for symmetric tensor representations related to $U_{q}(\widehat{sl_{n}})$.}

\author{Gary Bosnjak, Vladimir
  V.~Mangazeev}
\address{Department of Theoretical Physics,
         Research School of Physics and Engineering,\\
    Australian National University, Canberra, ACT 0200, Australia.}

\begin{abstract}
In this paper we construct a new factorized 
representation of  the $R$-matrix related to the affine algebra 
$U_{q}(\widehat{sl_{n}})$
for symmetric tensor representations with arbitrary weights.
Using the 3D approach we obtain
explicit formulas for the matrix elements of the $R$-matrix
and give a simple proof that a ``twisted'' $R$-matrix is stochastic.
We also discuss symmetries of the $R$-matrix, its  degenerations
and compare our formulas with other results available in the literature.
\end{abstract}
\end{center}

\section{Introduction}

Originally quantum groups were discovered in the context of quantum integrable systems and
quantum inverse scattering method
(see, for example, \cite{Faddeev:1979,KRS81,KR1981} and \cite{Fad95} for a  review).
They were formally introduced by Drinfeld and Jimbo
\cite{Drinfeld86,Jim85} as certain deformations of universal enveloping
algebras of simple Lie algebras. The $R$-matrix as a solution of the Yang-Baxter
equation \cite{Bazhanov87,Jimbo86} plays the central role in the theory of quantum groups.

From the physical point of view different representations  of
quantum groups
allow us to construct different spaces of physical states of quantum integrable models.
The $R$-matrix becomes a linear operator acting in the
tensor product of two arbitrary representations.
In this paper we address the problem of finding explicit matrix elements
for the $R$-matrix ${R}(\lambda)$ associated with symmetric tensor representations of
the affine quantum algebra $U_{q}(\widehat{sl_{n}})$. The parameter $\lambda$ here plays the role
of a spectral parameter entering the evaluation homomorphism \cite{Jim85}.

The problem of calculating quantum $R$-matrices related to the highest weight representations
of the $U_q(\widehat{sl_n})$ algebra have been considered  by many authors.
The two most known methods are a fusion procedure \cite{Str79,KRS81,KR87a}
and the method of spectral decomposition \cite{Jim85,Del94}.
Another method is based on the explicit evaluation of the universal $R$-matrix \cite{TK92}
in the tensor product of two highest weight representations but this method is technically
challenging and has been successfully applied only for low rank algebras \cite{BOOS2010,Boos11}.
We also mention the approach of \cite{CDS16} where the calculation of the higher-spin
$sl(2)$ $R$-matrices
is based on factorization properties of the $L$-operator.

In this paper we use the 3D approach developed in \cite{Bazhanov:2005as,Bazhanov:2008rd,MBS13}
and apply it to the case of symmetric
representations of the $U_q(\widehat{sl_n})$ algebra.
Previously 
a closed formula for the higher-spin $R$-matrix of the 6-vertex model was obtained in \cite{Man14}
based on the positive solution of the tetrahedron equation  \cite{MBS13}. 
It can be interpreted as the $R$-matrix of
the higher-spin stochastic 6-vertex model
\cite{BCG16,Bor14,CP16,BP16}. Under a special choice of the spectral parameter \cite{Bor14} this model
degenerates into the $q$-Hahn system which corresponds to the most general ``chipping model''
introduced by Povolotsky \cite{Povol13}. Let us notice that the action of the $Q$-operator for the higher
spin 6-vertex model (see Section 6  in \cite{Man14b}) can be identified with the transition matrix
of the  Povolotsky's chipping model in \cite{Povol13}.

In the recent paper \cite{KMMO16} the above higher spin stochastic 6-vertex model
has been generalized to the case of symmetric representations of the higher rank
 $U_q(A_n^{(1)})$ (or $U_q(\widehat{sl_{n+1}})$).
 It was shown that even for a general $n$ the corresponding $R$-matrix
 satisfies the sum rule required for a stochastic interpretation.
 At a special point it gives a $n$ species generalization of the Povolotsky model.
 However, most of the results in \cite{KMMO16}
were obtained using the machinery of quantum groups. Our strategy
is to derive explicit formulas for the $R$-matrix related to symmetric
representations of $U_q(A_n^{(1)})$ extending the method of \cite{Man14}.

The structure of the paper is as follows.
In Section 2 we introduce the Boltzmann weights of the $3D$ model from \cite{MBS13} and 
a definition of a composite weight.
In Section 3 we consider the $n$-layer projection of the $3D$ model and obtain
the formula for the $R$-matrix in the form of the $(n-1)$-tuple sum.  For $n=2$ it corresponds
to the formula from \cite{Man14} up to a certain transformation.
In Section 4 we discuss symmetries  of the $R$-matrix.
In Section 5 we consider degenerations and derive a factorization formula
for the $R$-matrix. In Section 6 we compare our formulas with other results
available in the literature.  
In Section 7 we introduce a stochastic $R$-matrix and give a simple proof of a sum rule.
We also consider the corresponding $L$-operator and show that it is equivalent
to the $L$-operator from \cite{GGW2016}. In Section 8 we discuss the results and
Appendix A contains notations and some formulas used in the main text.

\section{The 3D Integrable Model}

In this section we recall a definition of the 3D integrable model with positive Boltzmann weights introduced
in \cite{MBS13}. The Boltzmann weights of the model are constructed from matrix elements of an operator
$\textbf{R}$ acting in the tensor product of three   Fock spaces $\mathcal{F}\otimes
\mathcal{F} \otimes \mathcal{F}$ with the orthonormal
basis $\ket{n_{1},n_{2},n_{3}} = \ket{n_{1}}\otimes \ket{n_{2}}
\otimes \ket{n_{3}}$,\quad $n_{i}\in \Z_{\geq0}$. The operator $\textbf{R}$
solves the tetrahedron equation
\begin{align}\label{tetraeqn}
\textbf{R}_{123}\textbf{R}_{145}\textbf{R}_{246}\textbf{R}_{356}=
\textbf{R}_{356}\textbf{R}_{246}\textbf{R}_{145}\textbf{R}_{123}.
\end{align}

With respect to the
 basis $\ket{n_{1},n_{2},n_{3}}$, the operator \textbf{R} has the following matrix elements
\begin{align}\label{Relements}
\mathbb{R}_{n_{1},n_{2},n_{3}}^{n'_{1},n'_{2},n'_{3}} = \braket{n_{1},n_{2},n_{3}|
\textbf{R} |n'_{1},n'_{2},n'_{3}}
\end{align}
with
\begin{align}\label{3DR}
\mathbb{R}_{n_{1},n_{2},n_{3}}^{n'_{1},n'_{2},n'_{3}} &=
\delta_{n_{1}+n_{2},n_{1}'+n_{2}'}\delta_{n_{2}+n_{3},n_{2}'+n_{3}'}
q^{-n_{2}(1+n_{1}+n_{3})-n'_{1}n'_{3}} \notag \\
&\times {n_{1}+n_{2} \brack n_{1}}_{q^{2}}\>
\phiss\left( \left.\begin{matrix} q^{-2n_{2}},\>q^{-2n'_{1}} \\  q^{-2n_{1}-2n_{2}}\quad\>\>\end{matrix}
 \right| q^{2},q^{2(1+n'_{3})}\right),
\end{align}
where we used standard notations for $q$-series from Appendix A and
$\phiss$ is a basic hypergeometric series defined in \eqref{A4}.

We shall refer to the matrix representation of the operator $\textbf{R}$ as the 3D $R$-matrix.
It is easy to write a matrix realization of the operator equation \eqref{tetraeqn}
in terms of matrix elements $\mathbb{R}_{n_{1},n_{2},n_{3}}^{n'_{1},n'_{2},n'_{3}}$.

Note that in \eqref{3DR} we used a slightly different presentation of \eqref{3DR} compared to \cite{MBS13}.
The two presentations are equivalent
and related by a change $r:= n_{2}-r$ in the summation variable.
We also notice that due to conservation laws we always have
$n_2\leq n_1+n_2$ and $n_1'\leq n_1+n_2$, so the hypergeometric function in \eqref{3DR}
doesn't require a regularization. All nonzero elements in \eqref{3DR} are positive
for $0<q<1$ as explained in \cite{MBS13}.

The R-Matrix \eqref{3DR} possesses a number of
symmetries which are generated by two elementary ones \beq
\mathbb{R}_{n_{1},n_{2},n_{3}}^{n'_{1},n'_{2},n'_{3}}=
\mathbb{R}_{n_{3},n_{2},n_{1}}^{n'_{3},n'_{2},n'_{1}},\quad
\mathbb{R}_{n_{1},n_{2},n_{3}}^{n'_{1},n'_{2},n'_{3}}=
q^{n_{3}-n_2+n_{1}^2-n_1'^2}\frac{(q^{2};q^{2})_{n'_{1}}
}{(q^{2};q^{2})_{n_{1}}}\mathbb{R}_{n'_{1},n_{3},n_{2}}^{n_{1},n'_{3},n'_{2}}.\\
\label{3Dsym} \eeq

They can be proved by using Heine's transformations of $\phiss$ series \eqref{Heine}.
We list here two other useful symmetries which follow from \eqref{3Dsym} \beq
\mathbb{R}_{n_{1},n_{2},n_{3}}^{n'_{1},n'_{2},n'_{3}} =
q^{n_{1}-n_2+n_{3}^2-n_3'^2}\frac{(q^{2};q^{2})_{n'_{3}}
}{(q^{2};q^{2})_{n_{3}}}\mathbb{R}_{n_{2},n_{1},n'_{3}}^{n'_{2},n'_{1},n_{3}}\label{3Dsym1} \eeq
and
\beq \label{3Dsym2}
\mathbb{R}_{n_{1},n_{2},n_{3}}^{n'_{1},n'_{2},n'_{3}}= q^{(n_{3}+n'_{3}+2n'_{1}-2n_{2}+1)(n_{1}-n'_{1})}
\prod_{i=1}^3\frac{(q^{2};q^{2})_{n'_{i}}
}{(q^{2};q^{2})_{n_{i}}}\>\mathbb{R}_{n'_{1},n'_{2},n'_{3}}^{n_{1},n_{2},n_{3}}. \eeq

Let us notice that up to the factor $q^{-n_{2}(1+n_{1}+n_{3})-n'_{1}n'_{3}}$
the expression \eqref{3DR} is a polynomial in $q^{2n'_3}$
and can be formally continued to negative values $n_3,n_3'<0$. So let us assume that
$n_i, n_i'\geq0$, $i=1,2$ and $n_3,n_3'\in\mathbb{Z}$ provided that all indices are still constrained
by delta-functions entering \eqref{3DR}. Then it is easy to find a transformation of
matrix elements of the 3D $R$-matrix under the replacement $q\to q^{-1}$
\beq
\left.\mathbb{R}_{n_{1},n_{2},n_{3}}^{n'_{1},n'_{2},n'_{3}}\right|_{q\to q^{-1}}
=q^{(n_1-n_2')(n_2-n_2'-1)}
\mathbb{R}_{n_{1},n_{2},-n_3'-1}^{n'_{1},n'_{2},-n_3-1}.\label{symq}
\eeq

Following \cite{MBS13} we define a composite weight
$\mathbb{S}_{\bs{i},\bs{j}}^{\bs{i'},\bs{j'}}(w)$
\begin{align}\label{CompWeightN}
\mathbb{S}_{\bs{i},\bs{j}}^{\bs{i'},\bs{j}'}(w) = \sum_{\bs{k}}w^{k_{1}}
\prod_{s=1}^{n} \mathbb{R}_{j_{s},i_{s},k_{s}}^{j_{s}',i_{s}',k_{s+1}},
\end{align}
where $\bs{i} = \{i_{1},i_{2},\dots, i_{n}\},\>\bs{j} = \{j_{1},j_{2},\dots,j_{n}\}$, etc.
and $k_{n+1} = k_{1}$.

This composite weight has a number of important properties. Firstly, the presence of delta functions in
\eqref{3DR} imply global conservation laws , namely
\begin{align}\label{globalcons1}
I=I',&\quad J=J',\\
I:=\sum_{n=1}^{N} i_{n},\quad I':=\sum_{n=1}^{N} i'_{n},&\quad J:=
\sum_{n=1}^{N} j_{n},\quad J':=\sum_{n=1}^{N} j'_{n},\label{globalcons2}
\end{align}
and therefore the matrix $\mathbb{S}$ with entries \eqref{CompWeightN} will have a block-diagonal form
indexed by integers $I,J = 0,\dots,\infty$. Secondly, standard arguments relating the tetrahedron and
Yang-Baxter equations imply that \eqref{CompWeightN} satisfies the Yang-Baxter equation
\begin{align}\label{YBEcomp}
\sum_{\bs{i'},\bs{j'}\bs{k'}}
\mathbb{S}_{\bs{i},\bs{j}}^{\bs{i'},\bs{j'}}(w)
\mathbb{S}_{\bs{i'},\bs{k'}}^{\bs{i''},\bs{k'}}(w')
\mathbb{S}_{\bs{j'},\bs{k'}}^{\bs{j''},\bs{k''}}(w'/w) =
\sum_{\bs{i'},\bs{j'}\bs{k'}}
\mathbb{S}_{\bs{j},\bs{k}}^{\bs{j'},\bs{k'}}(w'/w)
\mathbb{S}_{\bs{i},\bs{k'}}^{\bs{i'},\bs{k''}}(w')
\mathbb{S}_{\bs{i'},\bs{j'}}^{\bs{i''},\bs{j''}}(w),
\end{align}
and hence defines the $R$-Matrix \cite{Bazhanov:2005as}.
This $R$-Matrix is composite in the sense that it is a direct sum of ``smaller'' $R$-
matrices. It is thr fact which follows from considering the conservation laws \eqref{globalcons1} applied to
each component in \eqref{YBEcomp} and noticing that the equation reduces to a tensor sum of
an infinite number of the Yang-Baxter
equations on subspaces indexed by global parameters $I,J,K = 0 ,\dots,\infty$ defined
in \eqref{globalcons2}.

 In particular, it was argued in \cite{Bazhanov:2005as,MBS13}
that the subspace for each parameter $I$ is in fact the underlying space of the rank $I$
symmetric tensor representation of $U_{q}(\widehat{\mathfrak{sl}_{n}})$ and the action of $\mathbb{S}$ on
this space is the corresponding R-matrix
\begin{align}\label{Blockform}
\mathbb{S}_{\bs{i},\bs{j}}^{\bs{i'},\bs{j}'}(w) =
\bigoplus_{I,J=0}^{\infty}R^{(n)}_{I,J}(w).
\end{align}

The case $n=2$ in \eqref{CompWeightN} was considered in \cite{Man14} which resulted in a new formula for the
matrix elements of the R-Matrix for $U_{q}(\widehat{\mathfrak{sl}_{2}})$ acting in the tensor product of
representations of highest weight $I$ and $J$. Setting $I=J=1$ the formula gives  the R-matrix
for the 6-vertex model.

\section{The n-layer projection}

In this section we will generalize the result of
\cite{Man14} by taking the $n$-layer projection and construct
a new formula for the $U_{q}(\widehat{\mathfrak{sl}_{n}})$ $R$-matrix acting in the tensor product
of representations with weights $I\omega_{1}$ and $J\omega_{1}$ respectively.

First let us introduce some vector notations.
We denote by $\bs{i} := \{i_{1},\dots,i_{r}\}$
a set of positive integers $i_k\in\mathbb{Z}_{\geq0}$ with $r$ components and define
\beq
|\bs{i}| =\sum_{s=1}^{r}i_{s},\quad (\bs{i},\bs{j}) =
\sum_{s=1}^{r}i_{s}j_{s}.\label{sums1}
\eeq
Addition is done component-wise and we introduce two permutations
$\sigma$ and $\tau$ acting on $\bs{k}$ as
\begin{align}\label{permutations}
\sigma\{k_{1},\dots,k_{r}\} = \{k_{2},\dots,k_{r},k_{1}\}\\
\tau\{k_{1},\dots,k_{r}\} = \{k_{r},k_{r-1},\dots,k_{1}\}
\end{align}
of the vector coordinates. The dimension $r$ can take values $n$ and $n-1$ as explained below.

The Kronecker delta function of two vectors is zero unless all their components match, i.e.
\beq
\delta_{\bs{i},\bs{j}}=\prod_{s=1}^r \delta_{i_s,j_s}.\label{Kron}
\eeq

We also note that in discussing $\mathbb{S}(w)$ and $R^{(n)}_{I,J}(w)$
the vectors $\bs{i},\bs{j},\bs{i'},\bs{j'}$ have different dimensions.
When we use $\mathbb{S}(w)$, the $n$-layer composite weight, it is implied that
the dimension $r=n$. When we derive the expression for the R-Matrix $R^{(n)}_{I,J}$,
 it is implied
that the dimension $r=n-1$ because by fixing $I,J$ the relation \eqref{globalcons2}
implies that we can remove one of the indices. Typically we choose to remove last components
$i_{n},j_{n},i'_{n},j'_{n}$ and replace them with $I - |\bs{i}|, J-|\bs{j}|$
etc. except in certain cases 
 where it is more convenient to keep them.
Of course, in evaluating final expressions  the replacement
has to be made regardless.

Combining \eqref{3DR} and \eqref{CompWeightN} the composite weight
$\mathbb{S}_{\bs{i},\bs{j}}^{\bs{i'},\bs{j'}}(w)$  can be written as
\begin{align}\label{CompWeightSub}
\mathbb{S}_{\bs{i},\bs{j}}^{\bs{i'},\bs{j}'}(w) &=
\delta_{\bs{i}+\bs{j},\bs{i'}+\bs{j'}} \sum_{\bs{k}\in\mathbb{Z}^n_+}
\delta_{\bs{i}+\bs{k},\bs{i'}+\sigma\bs{k}}\>w^{k_{1}}
q^{-|\bs{i}|-(\bs{i},\bs{j})-(\bs{k},\bs{i}+\sigma^{-1}\bs{j'})}
\prod_{s=1}^{n} { i_{s} + j_{s} \brack i_{s} }_{q^{2}}\\
&\times \sum_{\bs{m}\in\mathbb{Z}^n_+}\prod_{s=1}^{n} \frac{(q^{-2i_{s}},q^{-2j'_{s}};q^{2})_{m_{s}}}
{(q^{2},q^{-2i_{s}-2j_{s}};q^{2})_{m_{s}}} q^{2|\bs{m}|+2(\bs{m},\sigma\bs{k})}. \notag
\end{align}
The above formula  contains $2n$ summations. The $n$ summations in
$\bs{k}$ are infinite ranging from 0 to $\infty$. The $n$ summations in $\bs{m}$
are restricted by $0\leq m_s\leq \min(i_s,j_s')$, $s=1,\ldots,n$
due to the presence of  Pochhammers symbols in the numerator.
Let us also notice that all sums in $m_s'$ truncate before
the Pochhammer symbols in the denominator become zero. Therefore,
there is no need for a regularization.

This formula is quite easy to simplify. The presence of delta functions in
$\eqref{CompWeightSub}$ lead to the following global conservation laws for
the spin indices $\bs{i},\bs{j},\bs{i'},\bs{j'}$,
\begin{align} \label{ConsLaws}
i_{1}+\dotsb+i_{n} = i_{1}'+\dotsb+i_{n}' = I,\quad j_{1}+\dotsb+j_{n} = j_{1}'+\dotsb+j_{n}' = J
\end{align}
which allows us to remove one of the indices from $\bs{i},\bs{j},\bs{i'},\bs{j'}$
once we fix integers $I,J$. Furthermore,  we can also express $k_{2},\dots,k_{n}$ in
terms of $k_{1}$ by the relations
\beq
k_{s+1} = k_{s} + i_{s} - i'_{s},\quad
k_{n} = k_{1} + \sum_{s=1}^{n-1}(i_{s}-i'_{s}) \label{k1knrel},
\eeq
which allows us to rewrite the sum in $\bs{k}$ as a single sum in $k_{1}$.
However, some care must be taken in computing this sum. Note that when $i'_{s}>i_{s}$ for
some $s$, the summation range of $k_{s}$ implies contributions to the sum for negative values of $k_{s+1}$ not
included in the expression \eqref{CompWeightSub}. These contributions turn out to be trivial.
To see that we first notice that
\begin{align}
\mathbb{R}_{n_{1},n_{2},n_{3}}^{n'_{1},n'_{2},n'_{3}} = 0,\quad n_{3} < 0, \quad n'_{3}\geq0\label{proj1}
\end{align}
This follows from \eqref{3Dsym1}  since the factor $1/(q^2;q^2)_{n_3}$ becomes zero
and all other factors are nonzero. Now let us look at the product in \eqref{CompWeightN} and assume
that there are contributions from negative values for some $k_s$, $s=1,\ldots,n$.
All $k_s$ can not be negative,
since $k_1\geq0$. Since the product is cyclic, we will always find at least one factor
$\mathbb{R}_{j_{s},i_{s},k_{s}}^{j_{s}',i_{s}',k_{s+1}}$ such that $k_s<0$ and $k_{s+1}\geq0$.
This factor will be equal to zero because of \eqref{proj1}. Therefore, all factors which contain some
negative $k_s$ automatically disappear and we can safely sum over $k_1$ from $0$ to $\infty$ in
\eqref{CompWeightSub} with substitutions \eqref{k1knrel}.
As one can easily see the sum on $k_1$ becomes a geometric series which converges
provided
\begin{align}\label{convcriteria}
wq^{-I-J}<1,\quad 0<q<1.
\end{align} 

Once this condition is satisfied for $w=\lambda^2>0$, the sum in \eqref{CompWeightN} has all positive terms,
since all matrix elements of the $3D$ $R$-matrix \eqref{3DR} are positive. Restricting
the result to fixed positive values $I$, $J$
we get the expression for matrix elements of the operator $R^{(n)}_{I,J}(w)$
in \eqref{Blockform}. The result reads
\begin{align}\label{SlSum}
\left[R_{I,J}^{{(n)}}
(\lambda)\right]_{\bs{i},\bs{j}}^{\bs{i'},\bs{j'}} =
\delta_{\bs{i}+\bs{j},\bs{i'}+\bs{j'}}\>
q^{\Psi}
 \prod_{s=1}^{n} {i_{s}+j_{s} \brack i_{s}}_{q^2}  \sum_{\bs{m}\in\mathbb{Z}^n_+}
\frac{q^{2|\bs{m}| + 2\sum\limits_{k\geq l}m_{k}(i_{l}-i'_{l})}}
{1-\lambda^{2}q^{2|\bs{m}|-I-J}} \prod_{s=1}^{n}
\frac{(q^{-2i_{s}},q^{-2j'_{s}};q^{2})_{m_{s}}}{(q^{2},q^{-2(i_{s}+j_{s})};q^{2})_{m_{s}}}
\end{align}
where
\beq\label{phase}
\Psi=
-2(\bs{i},\bs{j})+(\bs{i'},\bs{j'})-(I-|\bs{i}|)(J-|\bs{j}|)
+I(|\bs{i'}|-|\bs{i}|-1)+\sum\limits_{1\leq k<l\leq n-1}(i'_{k}j'_{l}-i_{k}j_{l})
\eeq
Here in the LHS of \eqref{SlSum} and in the expression for the phase factor \eqref{phase}
we used $(n-1)$-component  indices, see \eqref{sums1} with  $r=n-1$.
However, in the RHS of \eqref{SlSum} for compactness
we kept $n$-component external indices assuming that
we need to substitute $i_n,\>j_n,i'_n,\>j'_n$ from \eqref{ConsLaws}.
The formula has $n$ summation indices $m_{1},m_{2},\dots,m_{n}$
which truncate after finitely many terms.
Finally we notice that the sum $\sum_{k\geq l}$ in \eqref{SlSum}  taken over
$n\geq k\geq l\geq1$ can be restricted to the values $n-1\geq k\geq l\geq 1$, since it is equal to zero
for $k=n$ due to \eqref{ConsLaws}.

The case $n=2$ of \eqref{SlSum} was given in (75) of \cite{MBS13}.
This formula generates elements of a $\binom{I+n-1}{n-1}\times\binom{J+n-1}{n-1}$-dimensional
matrix determined by indices
$0\leq |\bs{i}|,|\bs{i'}| \leq I, 0\leq |\bs{j}|,|\bs{j'}| \leq J$.

As the next step we shall evaluate one sum in \eqref{SlSum} and reduce the total number of summations to
$n-1$. We use the same method as in \cite{Man14}.

We start with the
Lagrange interpolating formula
\begin{align}\label{Interpol}
\sum_{l=0}^{k} \frac{x}{x-q^{l}}\frac{q^{l}(q^{-k};q)_{l}}{(q;q)_{l}}P_{k}
(q^{l})=\frac{P_{k}(x)(q;q)_{k}}{x^{k}(x^{-1};q)_{k+1}},
\end{align}
which is valid for any polynomial $P_{k}(x)$ of degree at most $k$.
First we define  a new variable
\begin{align}
l = m_{1}+\dotsb+m_{n}
\end{align}
which runs from 0 to $I$ and use $l$ instead of $m_{n}$. Then one can
rewrite \eqref{SlSum}   as
\begin{align}\label{SlSum2}
&\left[R_{I,J}^{(n)}(\lambda)\right]_{\bs{i},
\bs{j}}^{\bs{i'},\bs{j'}} =
\delta_{\bs{i}+\bs{j},\bs{i'}+\bs{j'}}(-1)^{|\bs{i}|-|\bs{i'}|}
q^{-2(\bs{i},\bs{j})+(\bs{i'},\bs{j'})-(I-|\bs{i}|)
(J-|\bs{j}|)+I(|\bs{i'}|-|\bs{i}|-1)+\sum_{k<l}(i'_{k}j'_{l}-i_{k}j_{l})} \notag\\
&\times \frac{q^{(|\bs{i}|-|\bs{i'}|)(|\bs{j}|+|\bs{j'}|-2J-1)}
(q^{-2J};q^{2})_{|\bs{j}|}}{(q^{-2J};q^{2})_{|\bs{j'}|}(q^{2};q^{2})_{I}}
\prod_{s=1}^{n-1} {i_{s}+j_{s} \brack i_{s}}  \sum_{l=0}^{I}\frac{q^{2l}}
{1-\lambda^{2}q^{-I-J+2l}} \frac{(q^{-2I};q^{2})_{l}}{(q^{2};q^{2})_{l}}P(q^{2l}).
\end{align}
The summation in $l$ matches \eqref{Interpol} with
$k=I$, $x = \lambda^{-2}q^{I+J}$
and
\begin{align}\label{Sum2}
P_I(q^{2l}) &= q^{2l(I+|\bs{i}|-|\bs{i'}|)}\sum_{\bs{m}}
q^{2(|\bs{m}|+\sum_{l>k}m_{k}(i'_{l}-i_{l}))}\prod_{s=1}^{n-1}
\frac{(q^{-2i_{s}},q^{-2j'_{s}};q^{2})_{m_{s}}}{(q^{2},
q^{-2(i_{s}+j_{s})};q^{2})_{m_{s}}}\\
&\times (q^{-2l};q^{2})_{|\bs{m}|}(q^{2(1-l+J-|\bs{j'}|+
|\bs{m}|};q^{2})_{I-|\bs{i'}|}(q^{2(1-l+I-|\bs{i}|+
|\bs{m}|)};q^{2})_{|\bs{i}|-|\bs{m}|}. \notag
\end{align}
The polynomial $P_I(x)$ in \eqref{Sum2} has degree of at most
$I$ and therefore we can replace the sum in $l$ in \eqref{SlSum2} with the right hand side
of $\eqref{Interpol}$ to find the expression
\begin{align}\label{slnfinal}
&\left[R_{I,J}^{(n)}
(\lambda)\right]_{\bs{i},\bs{j}}^{\bs{i'},\bs{j'}}=
\delta_{\bs{i}+\bs{j},\bs{i'}+\bs{j'}}
\left[A_{I,J}^{(n)}
(\lambda)\right]_{\bs{i},\bs{j}}^{\bs{i'},\bs{j'}}B_{I,J}(\lambda)\>
 q^{(\bs{i'},\bs{j'})
-(\bs{i},\bs{j})-J|\bs{i}|-I|\bs{j'}|+
\sum\limits_{k>l}(i_{k}j_{l}+j'_{k}i'_{l})}  \\
&\times \sum_{\bs{m}\in\mathbb{Z}^{n-1}_+}  \frac{(\lambda^{2}q^{-I-J},\lambda^{2}
q^{2+I+J-2|\bs{i}|-2|\bs{j}|};q^{2})_{|\bs{m}|}}{(\lambda^{2}q^{2+I-J-2|\bs{i}|},\lambda^{2}
q^{2+J-I-2|\bs{j'}|};q^{2})_{|\bs{m}|}}\prod_{s=1}^{n-1} \frac{(q^{-2i_{s}},
q^{-2j'_{s}};q^{2})_{m_{s}}}{(q^{2},q^{-2(i_{s}+j_{s})};q^{2})_{m_{s}}}
q^{2(|\bs{m}|+\sum\limits_{k<l}m_{k}(i'_{l}-i_{l}))}. \notag
\end{align}
All external and summation indices in \eqref{slnfinal} have $n-1$ components and
 the coefficients
$A_{I,J}^{(n)}(\lambda)_{\bs{i},\bs{j}}^{\bs{i'},\bs{j'}}$
and $B_{I,J}(\lambda)$ are given by
\begin{align}\label{slncoeff}
\left[A_{I,J}^{(n)}
(\lambda)\right]_{\bs{i},\bs{j}}^{\bs{i'},\bs{j'}} &=
\frac{(\lambda^{-2}q^{I-J};q^{2})_{|\bs{j'}|}(\lambda^{-2}q^{J-I};q^{2})_{|\bs{i}|}
(q^{-2J};q^{2})_{|\bs{j}|}}{(\lambda^{-2}q^{-I-J};q^{2})_{|\bs{i}+\bs{j}|}
(q^{-2J};q^{2})_{|\bs{j'}|}} \prod_{s=1}^{n-1} { i_{s} + j_{s} \brack j_{s}}_{q^{2}},\\
B_{I,J}(\lambda) &= q^{-I-IJ} \frac{(\lambda^{2}q^{-I-J};q^{2})_{I+J+1}}
{(\lambda^{2}q^{-I-J};q^{2})_{I+1}(\lambda^{2}q^{-I-J};q^{2})_{J+1}}.\label{slncoeff1}
\end{align}
The formula \eqref{slnfinal} gives the answer
 for the matrix elements of the $U_{q}(\widehat{\mathfrak{sl}_{n}})$ $R$-Matrix acting
 on the space $V_{I}\otimes V_{J}$
 where  
 \beq\label{Vn}
 V_I\equiv\{|{\bs{i}}\rangle\},\quad |\bs{i}|\leq I.
 \eeq 
 It follows from
 the tetrahedron equation for the 3D $R$-matrix \eqref{3DR} that
  $\eqref{slnfinal}$ satisfies the Yang-Baxter equation
\begin{align}\label{YBE}
R_{I,J}^{(n)}(\lambda)R_{I,K}^{(n)}(\lambda\mu)R_{J,K}^{(n)}(\mu)=
R_{J,K}^{(n)}(\mu)R_{I,K}^{(n)}(\lambda\mu)R_{I,J}^{(n)}(\lambda)
\end{align}
for any $I,J,K \in \Z_{+}$.
However, one will notice that the coefficient $B_{I,J}(\lambda)$ is
just a constant  not depending on indices. We find it convenient
to set this factor to 1. In what follows, we will use \eqref{slnfinal} with
$B_{I,J}(\lambda)=1$ unless stated otherwise.
In this normalization we have
\beq
\left[R_{I,J}^{(n)}
(\lambda)\right]_{\bs{0},\bs{0}}^{\bs{0},\bs{0}}=1.\label{newnorm}
\eeq

The main reason we do this is because \eqref{slnfinal}
 is now well defined even when $I,J\in \C$.
Although the 3D model projection outlined in this paper satisfies the Yang-Baxter equation
for integral weights by construction, the equation \eqref{YBE} remains valid even for
complex weights $I,J,K\in\C$. The proof closely follows the arguments given in \cite{KMMO16}.

Consider  a particular matrix element of the Yang-Baxter equation
$\langle{\bf i,j,k}|\mbox{\eqref{YBE}}|\bf{i',j',k'}\rangle$
with fixed external indices ${\bf{i}}=(i_1,\ldots,i_{n-1})$, etc.
Due to the conservation law in \eqref{slnfinal} we have
$|{\bf{i+j+k}}|=|{\bf{i'+j'+k'}}|\equiv
m$ and all summation indices in \eqref{YBE} will also be limited by $m$.
Choose an integer $N>m$ and assume that integer weights $I,J,K>N$.
It is easy to see that  all denominators in the $R$-matrices entering the Yang-Baxter
equation are non-zero and \eqref{YBE} becomes the equality of two rational functions
in variables $x=q^{-I}$, $y=q^{-J}$ and $z=q^{-K}$.
After eliminating denominators
we can rewrite \eqref{YBE} as equality of two polynomials in three variables
$x,y,z$. The degree of these polynomials grows as a fixed polynomial in $N$.
Now we know that the Yang-Baxter equation \eqref{YBE} is true for
infinitely many integer variables $I,J,K>N$.
It can only happen if \eqref{YBE} reduces to a polynomial identity in $x,y,z\in\C$.
Therefore, the Yang-Baxter equation with the $R$-matrix \eqref{slnfinal}
and normalization \eqref{newnorm} is satisfied for $I,J,K\in\C$.
In this case it defines the infinite dimensional $R$-matrix
 corresponding to Verma module representations
of $U_{q}(\widehat{\mathfrak{sl}_{n}})$.

To illustrate how formula \eqref{slnfinal} works, let us consider a special case $n=2$.
In this case matrix elements are indexed by indices $i,j,i',j'$,
\eqref{slnfinal} becomes a single sum which is given 
by 
\begin{align}\label{sl2}
\left[{R}_{I,J}^{(2)}(\lambda)\right]_{i,j}^{i',j'} &= \delta_{i+j,i'+j'}
q^{i'j'-ij-iJ-Ij'}{i+j \brack i}_{q^{2}} \frac{(\lambda^{-2}q^{I-J};q^{2})_{j'}
(\lambda^{-2}q^{J-I};q^{2})_{i}(q^{-2J};q^{2})_{j}}
{(\lambda^{-2}q^{-I-J};q^{2})_{i+j}(q^{-2J};q^{2})_{j'}}\notag\\
&\times{} _{4}\phi_{3}\left(
\left.
\begin{matrix}
q^{-2i}&q^{-2j'}&\lambda^{2}q^{-I-J}&\lambda^{2}q^{2+I+J-2i-2j} \\
 & q^{-2i-2j}&\lambda^{2}q^{2+I-J-2i}&\lambda^{2}q^{2+J-I-2j'}
\end{matrix}
\right| q^{2},q^{2}
\right).
\end{align}
This is a truncated and balanced  basic hypergeometric series $\phist$
for the elements of the $U_{q}(\widehat{\mathfrak{sl}_{2}})$ R-matrix.
This case was already studied in \cite{Man14} and the formula given there  is of
the same type as \eqref{sl2} but with different arguments.
Most notably, the hypergeometric sum in \cite{Man14} is a polynomial
in the spectral parameter $\lambda$ while \eqref{sl2} is a rational function.

Using the Sears transform \eqref{Sears} we can transform the sum in \eqref{sl2}
to (5.8) in \cite{Man14} by identifying $n=i,a=q^{-2j'}$, $b=\lambda^{2}q^{-I-J}$,
$c=\lambda^{2}q^{2+I+J-2i-2j}$, $d=\lambda^{2}q^{2+I-J-2i},e=q^{-2i-2j}$ and
$f=\lambda^{2}q^{2+J-I-2j'}$.
Let us note that  (5.8) in \cite{Man14} requires a regularization
but
the expression \eqref{sl2} is free from any divergences.

One of the problems  with \eqref{slnfinal} is that the hypergeometric sum
 is a rational function
in $\lambda$. Of course, being  multiplied with the extra factors \eqref{slncoeff}
it becomes a polynomial in $\lambda$.

As mentioned before, for $n=2$ \eqref{slnfinal} can be transformed to a polynomial formula in
$\lambda$ using the  Sears' transformation. The authors are aware of the $A_n$ multivariable
generalizations of the Sears' transformation in the literature, but they do not appear
to be applicable to our expression for $n\geq 3$. Focusing on each summation index
in \eqref{slnfinal} one can easily see that it is a ${}_{4}\phi_{3}$ basic
hypergeometric series but it is not balanced and only one of the hypergeometric series
 has a $q^{2}$  argument  so the $A_1$ Sears' transformations does not apply.

We expect that a formula with a hypergeometric sum being a polynomial in $\lambda$
still exists but it
 probably requires a new yet to be discovered identity
for multivariable hypergeometric series.

\section{Symmetries and special cases}

In this section we discuss  symmetries of the $R$-matrix $R_{I,J}^{(n)}(\lambda)$
given by \eqref{slnfinal}. They can be derived from the corresponding
symmetries of the 3D R-Matrix generated by \eqref{3Dsym}. It is actually more convenient
to use (\ref{3Dsym1}-\ref{3Dsym2}) since we need to to keep a position of the 3rd
``hidden'' direction where we take the trace.
Applying these transformations to the factors in \eqref{CompWeightN}
we find two symmetries
\begin{align}\label{2Dsym1}
\left[{R}_{I,J}^{(n)}
(\lambda)\right]_{\bs{i},\bs{j}}^{\bs{i'},\bs{j'}} &=
\lambda^{2(|\bs{i'}|-|\bs{i}|)}
\left[{R}_{J,I}^{(n)}
(\lambda)\right]_{\tau\bs{j},\tau\bs{i}}^{\tau\bs{j'},\tau\bs{i'}},
\\
\left[{R}_{I,J}^{(n)}(\lambda)
\right]_{\bs{i},\bs{j}}^{\bs{i'},\bs{j'}} &=
q^{2[\bs{i'},\bs{j'}]-2[\bs{i},\bs{j}]}
\lambda^{2(|\bs{i'}|-|\bs{i}|)}\prod_{s=1}^{n}
\frac{(q^{2};q^{2})_{i'_{s}}(q^{2};q^{2})_{j'_{s}}}
{(q^{2};q^{2})_{i_{s}}(q^{2};q^{2})_{j_{s}}}\left[
{R}_{I,J}^{(n)}(\lambda)
\right]_{\tau\bs{i'},\tau\bs{j'}}^{\tau\bs{i},\tau\bs{j}}.\label{2Dsym2}
\end{align}
Let us explain some notations here. In the previous section we mentioned
that for the $R$-matrix $R_{I,J}^{(n)}(\lambda)$ we are using $n-1$-component
indices, i.e.  $\bs{i}=\{i_1,\ldots,i_{n-1}\}$ with  the last $n$-th component
$i_n=I-|\bs{i}|$ removed and similar for $j$'s. However, in \eqref{2Dsym2}
the product in the RHS is taken over $s=1,\ldots,n$ where for $s=n$ we substitute
the last component as above, i.e. $i_n=I-|\bs{i}|$, $j_n=J-|\bs{j}|$, etc.
The transformation $\tau$ is defined in \eqref{permutations}.

In addition, in \eqref{2Dsym2} we used a notation $[\bs{i},\bs{j}]$ for
a convolution of $n$-component indices, i.e.
\beq
[\bs{i},\bs{j}]=(\bs{i},\bs{j})+(I-|\bs{i}|)(J-|\bs{j}|).\label{extra1}
\eeq

There is also a symmetry of the $R$-matrix which corresponds the the cyclic permutation of
the $n$ $3D$ $R$-matrices in the ``hidden'' direction.
Let us introduce the notation
\begin{align}
\bar{\bs{i}} = \{I-|\bs{i}|,i_{1},\dots,i_{n-2}\}.\label{extra2}
\end{align}
which is equivalent to the permutation $\sigma^{-1}\bs{i}$ for the $n$-tuple
$\bs{i}$ but with the last component removed.
Here we assume that $I,J\in\mathbb{Z}_+$.
Performing a cyclic shift in \eqref{CompWeightN} we easily obtain
\beq
\left[{R}_{I,J}^{(n)}(\lambda)
\right]_{\bs{i},\bs{j}}^{\bs{i'},\bs{j'}} =
\lambda^{2(|\bs{i'}|-|\bs{i}|)}\left[
{R}_{I,J}^{(n)}(\lambda)
\right]_{\bar{\bs{i}},\bar{\bs{j}}}^{\bar{\bs{i}'},\bar{\bs{j'}}}.
\eeq
For example, when $n=2$ this corresponds to $i\to I-i$ and similarly for other indices.
\\

The last symmetry follows from  the transformation of the $3D$ $R$-matrix \eqref{symq}.
After simple calculations one can obtain the following
result
\begin{align}
\left[{R}_{I,J}^{(n)}(\lambda,q)
\right]_{\bs{i},\bs{j}}^{\bs{i'},\bs{j'}} &=
q^{[\bs{i},\bs{j}]-[\bs{i'},\bs{j'}]}
\left[{R}_{J,I}^{(n)}(\lambda^{-1},q^{-1})
\right]_{\bs{j},\bs{i}}^{\bs{j'},\bs{i'}}.
\end{align}

Finally, when $I=J$ and $\lambda=1$ the $R$-matrix reduces to permutation operator
\beq
{R}_{I,I}^{(n)}(1)=\mathcal{P}_{1,2} \label{perm}
\eeq
which can be seen from \eqref{degen2} in the next section.

\section{Reductions and factorization}

There are two special points in the spectral parameter $\lambda=q^{\pm(I-J)/2}$
where the multiple sum in \eqref{slnfinal}
reduces to one non-zero summand.
These specializations produce the $R$-matrix without difference property
with weights $I,J$ playing the role of  spectral parameters.
With the normalization \eqref{newnorm} we can choose $I,J\in\mathbb{C}$ and
obtain the $R$-matrix acting in the tensor product of two Verma modules.

For the case of the $U_q(\widehat{sl_2})$ algebra the importance of such reductions
was first noticed in \cite{Bor14}. Under the choice $\lambda=q^{(J-I)/2}$
the $U_q(\widehat{sl_2})$ $R$-matrix of \cite{Man14} reduces to the
$R$-matrix of Povolotsky model \cite{Povol13} which satisfies stochasticity
condition and defines a family of zero-range chipping models.
A generalization of the Povolotsky model to arbitrary rank $n-1$ was
 obtained
in the recent paper \cite{KMMO16}.

So let us start with the case $\lambda = q^{(I-J)/{2}}$, $I-J\in\mathbb{Z}_+$.
The expression for the $R$-matrix
\eqref{slnfinal} contains the factor $(\lambda^{-2}q^{I-J};q^{2})_{|\bs{j'}|}$
outside the sum which has the argument $1$ after the above substitution.
This factor is always zero for $|\bs{j'}|>0$
unless it is canceled off by the factor $(\lambda^{2}q^{2+J-I-|\bs{j'}|};q^{2})_{|\bs{m}|}$
inside the sum. It can only happen when $|\bs{m}|=|\bs{j'}|$ or $\bs{m}=\bs{j'}$,
since $m_s\leq j'_s$, $s=1,\ldots,n-1$.
Let us note that the argument fails when $J-I$ is a positive integer because 
the other factor in the denominator
can cancel off the zero of $(\lambda^{-2}q^{I-J};q^{2})_{|\bs{j'}|}$ and multiple summands survive.

After simple algebra one can derive from \eqref{slnfinal} the following result
\begin{align}
\left[{R}_{I,J}^{(n)}(q^{\frac{I-J}{2}})
\right]_{\bs{i},\bs{j}}^{\bs{i'},\bs{j'}} &=
\delta_{\bs{i}+\bs{j},\bs{i'}+\bs{j'}}
q^{(\bs{i'},\bs{j'})-(\bs{i},\bs{j})-J|\bs{i}|-I|
\bs{j'}|+2J|\bs{j'}|+\sum\limits_{k>l}(i_{k}j_{l}+j'_{k}i'_{l}-2j'_{k}j_{l})
} \notag \\
&\times \frac{(q^{-2J};q^{2})_{|\bs{j}|}(q^{2J-2I};
q^{2})_{|\bs{i'}|-|\bs{j}|}}{(q^{-2I};q^{2})_{|\bs{i'}|}}
\prod_{s=1}^{n-1} { i'_{s} \brack j_{s}}_{q^{2}}.\label{degen1}
\end{align}

Similarly, we can  make substitution $\lambda = q^{\frac{J-I}{2}}$ for $J-I\in\mathbb{Z}_+$.
In this case the argument is the same except with the factors
$(\lambda^{-2}q^{J-I};q^{2})_{|\bs{i}|}$ and
$(\lambda^{2}q^{2+I-J-|\bs{i}|};q^{2})_{|\bs{m}|}$ and so
the only summand that contributes is $\bs{m}=\bs{i}$. Then we obtain
\begin{align}
\left[{R}_{I,J}^{{(n)}}(q^{\frac{J-I}{2}})
\right]_{\bs{i},\bs{j}}^{\bs{i'},\bs{j'}} &=
\delta_{\bs{i}+\bs{j},\bs{i'}+\bs{j'}}
q^{(\bs{i'},\bs{j'})-(\bs{i},\bs{j})-J|
\bs{i}|-I|\bs{j'}|+2I|\bs{i}|+
\sum_{k>l}(i_{k}j_{l}+j'_{k}i'_{l}-2i_{k}i'_{l})}\notag \\
&\times \frac{(q^{-2I};q^{2})_{|\bs{i}|}(q^{2I-2J};
q^{2})_{|\bs{j'}|-|\bs{i}|}}{(q^{-2J};q^{2})_{|\bs{j'}|}}
\prod_{s=1}^{n-1} { j'_{s} \brack i_{s}}_{q^{2}}.\label{degen2}
\end{align}
Obviously these reductions  are substantially simpler than the
original $R$-matrix. As mentioned above $I,J$ play role of the spectral parameters
for these $R$-matrices and can now take arbitrary complex values.

In fact, one can construct the full $R$-matrix
as a matrix product of  \eqref{degen1}-\eqref{degen2}.
To explain this it is convenient to apply a simple similarity transformation
in the first space and introduce
\beq
\bs{R}_{I,J}^{(n)}(\lambda)= U\otimes \bs{1}\>{R}_{I,J}^{{(n)}}(\lambda)\>
U^{-1}\otimes \bs{1}\label{degen3}
\eeq
with
\beq\label{degen4}
U_{\bs{i},\bs{i'}}=\delta_{\bs{i},\bs{i'}}\left(\lambda q^{(I-J)/2}\right)^{|\bs{i}|}.
\eeq
Now let us define two operators $\bs{M}$ and $\bs{N}$ acting in the tensor product of two Verma modules
by
\beq
\bs{M}(q^{I},q^{J}) = \bs{\check R}_{I,J}^{(n)}(q^{\frac{I-J}{2}}),\quad
\bs{N}(q^{I},q^{J}) = \bs{\check R}_{I,J}^{(n)}(q^{\frac{J-I}{2}})\label{degen5b}
\eeq
where as usual $\check {R}_{1,2}(\lambda)=\mathbb{P}_{1,2} R_{1,2}(\lambda)$, etc.
with $\mathbb{P}_{1,2}$ being the permutation operator.
Both operators $\bs{M}(q^{I},q^{J})$ and
$\bs{N}(q^{I},q^{J})$
of complex arguments
$q^I$, $q^J$ are defined
by its matrix elements via \eqref{degen1}-\eqref{degen2} and \eqref{degen5b}.

With these notations one can easily derive from \eqref{slnfinal} the following factorization
\beq
\bs{\check R}_{I,J}^{(n)}(\lambda) =
\bs{M}(\lambda q^{\frac{I+J}{2}},q^{J})
\bs{N}(\lambda^{-1}q^{\frac{I+J}{2}},q^{J}).\label{degen6}
\eeq
A similar factorization of the $R$-matrix appeared in \cite{DerMan2010} for the $n=2$ case
of the XXX
chain.

We can also rewrite a factorization formula \eqref{degen6}
for the matrix elements of the original $R$-matrix
${R}_{I,J}^{{(n)}}(\lambda)$ as follows
\beq
\left[{R}_{I,J}^{(n)}(\lambda)\right]_{\bs{i},
\bs{j}}^{\bs{i'},\bs{j'}}
= \sum_{\bs{k}+\bs{l}=\bs{i}+\bs{j}}
\bs{\tilde M}_{\bs{i},\bs{j}}^{\bs{k},
\bs{l}}\>\bs{\tilde N}_{\bs{k},
\bs{l}}^{\bs{i'},\bs{j'}},\label{degen7}
\eeq
with
\begin{align}
&\bs{\tilde{M}}_{\bs{i},\bs{j}}^{\bs{i'},\bs{j'}} =
\delta_{\bs{i}+\bs{j},\bs{i'}+\bs{j'}}\>
q^{-J|\bs{i}|-(\bs{i},\bs{j})+
\sum\limits_{k>l}(i_{k}j_{l}+j'_{k}i'_{l}-2 j'_{k}j_{l})}\>
\frac{(q^{-2J};q^{2})_{|\bs{j}|}(\lambda^{-2}q^{J-I};
q^{2})_{|\bs{i'}-\bs{j}|}}{( \lambda^{2}q^{-I-J})^{|\bs{j'}|}
(\lambda^{-2}q^{-I-J};q^{2})_{|\bs{i'}|}}
\prod_{s=1}^{n-1} { i'_{s} \brack j_{s}}_{q^{2}},\label{degen8a}\\
&\bs{\tilde{N}}_{\bs{i},\bs{j}}^{\bs{i'},\bs{j'}} =
\delta_{\bs{i}+\bs{j},\bs{i'}+\bs{j'}} \>
q^{(\bs{i'},\bs{j'})-I|\bs{j'}|+\sum\limits_{k>l}(j_{k}i_{l}+j'_{k}i'_{l}
-2 j_{k}i'_{l})}\>
\frac{(\lambda^{2}q^{-I-J};q^{2})_{|\bs{j}|}(\lambda^{-2}q^{I-J};
q^{2})_{|\bs{j'}-\bs{j}|}}{(q^{-2J};q^{2})_{|\bs{j'}|}}
\prod_{s=1}^{n-1} { j'_{s} \brack j_{s}}_{q^{2}}\label{degen8b}
\end{align}
where we removed some gauge factors which cancel in the matrix product \eqref{degen7}.

\section{Comparison  with other results}

In this and next sections we will compare  \eqref{slnfinal} with some  other
presentations of the $U_q(\widehat{sl_n})$ related $R$-matrix given in the literature.
We will establish a connection with the standard
the $U_q(\widehat{sl_{n}})$ $L$-operator presented in \cite{KMMO16}
and also compare our results with some higher-spin examples of
the $U_{q}(\widehat{\slthree})$ $R$-Matrix. 

We start with some  remarks regarding the coefficient
$A_{I,J}^{(n)}(\lambda)$ in \eqref{slncoeff}.
For specific elements of the $R$-matrix the $q$-Pochhammer symbols are finite
as their  arguments  are integers.
If we want to derive the formula for the $L$-operator as
an $n\times n$-matrix with operator
entries acting in the 
Verma modules spanned by $|\bs{j}\rangle=|j_1,\ldots,j_{n-1}\rangle$
 we need to rewrite \eqref{slncoeff}
in the form suitable for abstract values of $\bs{j}$ indices.

This is achieved by a slight change of normalization of the $R$-matrix
\begin{align}\label{slrenorm}
\bar{R}_{I,J}^{(n)}(\lambda) &= \sigma_{I,J}(\lambda){R}_{I,J}^{(n)}(\lambda),
\end{align}
with
\beq \label{newcoeff}
\sigma_{I,J}(\lambda) = -\lambda^{-I}q^{\frac{I+J}{2}}
(\lambda^{2}q^{-I-J};q^{2})_{I+1}.
\eeq
We also restore a coefficient $B_{I,J}(\lambda)$ in \eqref{slncoeff1} and define
\beq
\bar{A}_{I,J}^{(n)}(\lambda) = 
\sigma_{I,J}(\lambda)A_{I,J}^{(n)}(\lambda)B_{I,J}(\lambda).
\eeq
After simple calculations we obtain
\beq
\bar{A}_{I,J}^{(n)}(\lambda)
=\frac{(\lambda^{-2}q^{-I-J+2|\bs{i}|+2|\bs{j}|};q^{2})_{I-|\bs{i'}|}
(\lambda^{-2}q^{J-I};q^{2})_{|\bs{i}|}}{(-1)^{I+1}\lambda^{-I}
q^{-\frac{I+J}{2}}(q^{-2J+2|\bs{j}|};q^{2})_{|\bs{i-i'}|}}
\prod_{s=1}^{n-1}\frac{(q^{2+2j_{s}};q^{2})_{i_{s}}}
{(q^{2};q^{2})_{i_{s}}}\label{coeffnewform}
\eeq
and this expression is a finite product for integer $\bs{i}$, $\bs{i'}$
and abstract values of $\bs{j}$'s.
Shortly speaking a change of normalization is equivalent to replacing
the product $A_{I,J}^{(n)}(\lambda)B_{I,J}(\lambda)$ in 
\eqref{slnfinal} with \eqref{coeffnewform}.
The sum in \eqref{slnfinal} is still finite because it 
truncates by integer values of $\bs{i}$'s.

It is easier to write down explicit formulas in original $n$-component notations.
Introduce $n$-component vectors
$\bs{e}_\alpha=(0,\ldots,0,1,0,\ldots,0)$ with $1$'s at
the $\alpha$-th position from the left,
$\bs{j}, \bs{k}\in\mathbb{Z}^n_+$ with $|\bs{j}|=|\bs{k}|=J$.
Then using \eqref{coeffnewform} in \eqref{slnfinal}
we obtain for the renormalized $R$-matrix \eqref{slrenorm}
\begin{align}\label{slnlop}
\left[\bar{R}^{(n)}_{1,J}(\lambda)
\right]_{\bs{e}_{\alpha},\bs{j}}^{\bs{e}_{\beta},
\bs{k}} =  \begin{cases}
     [\lambda q^{\frac{1-J}{2}+k_{\alpha}}] &
     \text{if}\ \alpha=\beta, \\
      \lambda q^{\frac{1-J}{2}+\sum_{s=\beta}^{\alpha-1}k_{s}}
      [q^{k_{\alpha}}]& \text{if} \ \alpha>\beta, \\
        \lambda^{-1} q^{\frac{1+J}{2}-
        \sum_{s=\alpha}^{\beta-1}k_{s}}[q^{k_{\alpha}}]& \text{if} \ \alpha<\beta,
   \end{cases}
\end{align}
where
\beq
[x]=x-x^{-1}.\label{bra}
\eeq
In the recent paper \cite{KMMO16} matrix elements for the
$U_{q}(A_{n-1}^{(1)})$ $R$-Matrix  $R^K(z)$ acting in the space
$V_{1} \otimes V_{m}$ were given by
\begin{align}\label{kunibar1}
\left[R^{K}_{1,m}(z)\right]_{e_{j},
\bs{\beta}}^{e_{k},\bs{\delta}} =  \begin{cases}
      q^{\beta_{k}+1}\frac{1-q^{-2\beta_{k}+m-1}z}
      {q^{m+1}-z} & \text{if}\ j=k \\
      -q^{\beta_{j+1}+\dots+\beta_{k-1}} \frac{1-q^{2\beta_{k}}}
      {q^{m+1}-z}& \text{if} \ j<k, \\
        -q^{m-(\beta_{k}+\dots+\beta_{j})} 
        \frac{z(1-q^{2\beta_{k}})}{q^{m+1}-z} & \text{if} \ j>k.
   \end{cases}
\end{align}
and the elements of $R^K(z)$ acting on $V_{l} \otimes V_{1}$ were given by
\begin{align}\label{kunibar2}
\left[R^{K}_{l,1}(z)\right]_{\bs{\alpha},
e_{j}}^{\bs{\gamma},e_{k}} =  \begin{cases}
      q^{\gamma_{k}+1}\frac{1-q^{-2
      \gamma_{k}+l-1}z}{q^{l+1}-z} & \text{if}\ j=k \\
   -q^{l-(\alpha_{j}+\dots+\alpha_{k})} 
   \frac{z(1-q^{2\alpha_{k}})}{q^{l+1}-z} & \text{if} \ j<k, \\
     -q^{\alpha_{k+1}+\dots+\alpha_{j-1}} 
     \frac{1-q^{2\alpha_{k}}}{q^{l+1}-z}  & \text{if} \ j>k,
   \end{cases}
\end{align}
where we write them in the same notations as in \eqref{slnlop}.
\\
A direct comparison of \eqref{kunibar1} and \eqref{slnlop} gives
\beq
\left[\bar{R}^{(n)}_{1,J}(\lambda)\right]_{\bs{e}_{\alpha},
\bs{j}}^{\bs{e}_{\beta},
\bs{k}}=
\left[\lambda q^{\frac{1+J}{2}}\right]q^{[\bs{e}_{\beta},
\bs{k}]-[\bs{e}_{\alpha},\bs{j}]}
\left[R^{K}_{1,J}(\lambda^{-2})\right]^{\bs{e}_{\alpha},
\bs{j}}_{\bs{e}_{\beta},
\bs{k}}.\label{compar1}
\eeq
To compare matrix elements of $R^K_{l,m}(z)$ with our 
formula $\eqref{slnfinal}$ for other cases
we must first identify their  parameters. 
So we set $l=I$, $m=J$ and $z=\lambda^{-2}$.
The normalization of the $R$-matrix $R^K_{l,m}(z)$ is the same as \eqref{newnorm}
for $R_{I,J}^{(n)}(\lambda)$. Therefore, we expect that for arbitrary $I,J$
\beq
\left[{R}^{(n)}_{I,J}(\lambda)\right]_{\bs{i},\bs{j}}^{\bs{i'},
\bs{j'}}=
q^{[\bs{i'},
\bs{j'}]-[\bs{i},\bs{j}]}
\left[R^{K}_{I,J}(\lambda^{-2})\right]^{\bs{i},\bs{j}}_{\bs{i'},
\bs{j'}}.\label{compar2}
\eeq
The difference between two $R$-matrices in \eqref{compar2} 
is easy to explain.  The matrix
elements of ${R}^{(n)}_{I,J}(\lambda)$ are defined similar to \eqref{Relements}, i.e.
\beq
\left[{R}^{(n)}_{I,J}(\lambda)\right]_{\bs{i},\bs{j}}^{\bs{i'},
\bs{j'}}=\langle \bs{i},\bs{j}|{R}^{(n)}_{I,J}(\lambda)|\bs{i'},
\bs{j'}\rangle. \label{compar3}
\eeq
However, the matrix elements of $R^{K}_{l,m}(z)$ in \cite{KMMO16} 
are defined by the transposed
action
\begin{align}
\left[R^{K}_{l,m}(z)\right]_{\alpha,\beta}^{\gamma,\delta} =
 \braket{\gamma,\delta| R^{K}_{l,m}(z) |\alpha,\beta}.\label{compar4}
\end{align}
It is easy to check that the extra ``twist'' factor $q^{[\bs{i'},
\bs{j'}]-[\bs{i},\bs{j}]}$ in \eqref{compar2} drops out
from the Yang-Baxter equation.

We have checked that the relation \eqref{compar2} holds for
the $U_{q}(A_{1}^{(1)})$ and
$U_{q}(A_{2}^{(1)})$  $R$-matrices for all cases given in Appendix A of \cite{KMMO16}.

It is also interesting to compare our reductions \eqref{degen1} 
and \eqref{degen2} with that
obtained in \cite{KMMO16}. In particular, we expect that 
the {\bf Theorem 2} in \cite{KMMO16}
\begin{align}\label{kunibareduction}
\left[R^{K}_{l,m}(q^{l-m})\right]_{\alpha,\beta}^{\gamma,\delta} &=
\delta_{\alpha+\beta,\gamma+\delta} q^{\psi} {m \brack l}_{q^{2}}^{-1}
\prod_{s=1}^{n+1}{\beta_{s} \brack \gamma_{s}}_{q^{2}}\\
\psi &= \sum_{1\leq s,t \leq n+1} \alpha_{s}(\beta_{t}-\gamma_{t}) +
\sum_{1\leq s,t \leq n+1} (\beta_{s}-\gamma_{s})\gamma_{t}
\end{align}
should correspond to the substitution $\lambda = q^{\frac{J-I}{2}}$ 
given by \eqref{degen2}.
A direct calculation shows the relation \eqref{compar2} also holds in this case.

Now let us turn to the $U_q({sl_n})$ $L$-operator.
When $J=1$ the expression $\eqref{slnlop}$ further reduces to
the trigonometric $n$-state $R$-Matrix
 \cite{Cherednik1980,Kulish:1980ii}. We shall also use
a  {\it twisted} version of this $R$-matrix   \cite{Perk:1981nb}
  which we give using notations of \cite{Bazhanov:1990qk}
\begin{align}\label{BKMSRmat}
{R}_{\alpha,\gamma}^{\beta,\delta}(\lambda) =
\delta_{\alpha,\beta}\delta_{\gamma,\delta}\delta_
{\alpha,\gamma}(q-1)(\lambda+\lambda^{-1}q^{-1})+
\delta_{\alpha,\beta}\delta_{\gamma,\delta}
\rho_{\alpha,\gamma}
(\lambda-\lambda^{-1})+\delta_{\alpha,\delta}
\delta_{\beta,\gamma}\sigma_{\alpha,\beta},
\end{align}
where
\begin{align}\label{sigdef}
\sigma_{\alpha,\beta}=  \begin{cases}
 0& \text{if} \ \alpha=\beta, \\
     (q-q^{-1})\lambda& \text{if} \ \alpha<\beta, \\
        (q-q^{-1})\lambda^{-1}& \text{if} \ \alpha>\beta
   \end{cases}
\end{align}
and $\rho_{\alpha,\beta}$ are nonzero complex parameters  such that
\beq
\rho_{\al,\al}=\rho_{\al,\be}\rho_{\be,\al}=1,\quad \al,\be=1,\ldots,n.\label{rhos}
\eeq
Setting all ${\rho}_{\al,\be}=1$
and taking convention that
all indices $\alpha,\beta,\gamma,\delta=1,\ldots,n$ in \eqref{BKMSRmat}
denote positions of $1$'s counted from the right, i.e.
$\alpha\equiv \bs{e}_{n-\alpha+1}$ we obtain that \eqref{BKMSRmat}
is equivalent to \eqref{slnlop} with $J=1$.

Setting $I=J=1$ in the Yang-Baxter equation \eqref{YBE} we obtain the $L$-operator algebra
\begin{align}\label{YBELOP}
R_{1,2}(\lambda/\mu)L_{1}(\lambda)L_{2}(\mu) = 
L_{2}(\mu)L_{1}(\lambda)R_{1,2}(\lambda/\mu),
\end{align}
where the $R_{1,2}(\lambda)$-matrix corresponds to 
the standard $U_q(A^{(1)}_{n-1})$ trigonometric
$R$-matrix \eqref{slnlop} with $J=1$.
The $L$-operators $L(\lambda)$ are identified with
$\bar{R}_{1,K}^{(n)}(\lambda)$ \eqref{slnlop} acting in 
the ``quantum'' space with the weight $K$.

To rewrite the $L$-operator  in algebraic notations let us introduce
Weil operators $X_k,Z_k$, $i=1,\ldots,n$ acting in the space of $n$-component
vectors $|\bs{j}\rangle$, $j_s\in\mathbb{Z}$,
$s=1,\ldots,n$ and their conjugates such that
\beq
Z_k |\bs{j}\rangle=q^{j_k}|\bs{j}\rangle,\quad
X_k |j_1,\ldots,j_n\rangle=|j_1,\ldots,j_k+1,\ldots,j_n\rangle, \label{Weil1}
\eeq
\beq
\langle\bs{j}|Z_k=q^{j_k}\langle\bs{j}|,\quad
\langle j_1,\ldots,j_n|X_k=\langle j_1,\ldots,j_k-1,\ldots,j_n|. \label{Weil2}
\eeq
They  satisfy the Weil algebra relations
\beq
Z_k X_l=q^{\delta_{k,l}}X_l,Z_k,\quad k,l=1,\ldots,n.\label{Weil3}
\eeq
We can now define the $L$-operator $L(\lambda)$ as an  $n\times n$ matrix
with operator entries such that
\beq
\langle\bs{j}| L_{\al,\be}(\lambda)|\bs{k}\rangle=
\left[\bar{R}^{(n)}_{1,J}(\lambda)
\right]_{\bs{e}_{\alpha},\bs{j}}^{\bs{e}_{\beta},\label{Weil4}
\bs{k}}.
\eeq
Using \eqref{Weil1}-\eqref{Weil3} we obtain
\begin{align}\label{slnWeil}
L_{\al,\be}(\mu) =  \begin{cases}
    \left[\mu Z_\al\right] &
     \text{if}\ \alpha=\beta, \\
      \mu\> X^{-1}_\al X_\be\left[Z_\al\right]\prod\limits_{s=\be}^{\al-1}Z_s
      & \text{if} \ \alpha>\beta, \\
        \mu^{-1}q X^{-1}_\al X_\be \left[Z_\al\right]\prod\limits_{s=\al}^{\be-1}Z_s^{-1}
       & \text{if} \ \alpha<\beta,
   \end{cases}
\end{align}
where we defined a rescaled spectral parameter
$\mu=\lambda q^{\frac{1-J}{2}}$ and for any vector 
$|\bs{j}\rangle$, $|\bs{j}|=J$
\beq
\mathcal{Z}|\bs{j}\rangle=q^{|\bs{j}|} |\bs{j}\rangle,\quad
\mathcal{Z}=\prod_{s=1}^n Z_s\label{Weil5}
\eeq
In fact, we can consider \eqref{slnWeil} as an operator solution of the algebra 
\eqref{YBELOP}
since a rescaling of the spectral parameter does not affect \eqref{YBELOP}.
The operator $\mathcal{Z}$ commutes with \eqref{slnWeil} and
all representations are characterized by its complex eigenvalue $q^J$.

\section{Stochastic $R$-matrix}

Let us define another $R$-matrix $S_{I,J}(\lambda)$ by
\begin{align}
\left[S_{I,J}(\lambda)
\right]_{\bs{i},\boldsymbol{j}}^{\bs{i'},\bs{j'}} = 
\rho_{\bs{i},\bs{j}}^{\bs{i'},\bs{j'}}
\left[{R}_{I,J}^{(n)}(\lambda)
\right]_{\bs{i},\bs{j}}^{\bs{i'},\bs{j'}},\quad
\bs{i},\bs{j},\bs{i'},\bs{j'}\in\mathbb{Z}_+^{n-1},
 \label{stoc1}
\end{align}
with
\beq 
\rho_{\bs{i},\bs{j}}^{\bs{i'},\bs{j'}}=
q^{[\bs{i},\bs{j}]-[\bs{i'},\bs{j'}]
+\sum\limits_{1\leq k<l\leq n}(j_{k}i_{l}-i'_{k}j'_{l})}=
q^{(\bs{i},\bs{j})-(\bs{i'},\bs{j'})-J|\bs{i}|+
I|\bs{j'}|+\sum\limits_{1\leq k<l<n}(j_{k}i_{l}-i'_{k}j'_{l})}.
\eeq

In \cite{KMMO16} $S_{I,J}(z)$ was given in terms of $R^K_{I,J}(z)$ with
$z=\lambda^{-2}$.
Here we defined  $S_{I,J}(\lambda)$ in terms of $R^{(n)}_{I,J}(\lambda)$
using the relation \eqref{compar2}.
Using quantum group arguments it was shown in \cite{KMMO16} that
\eqref{stoc1} solves the Yang-Baxter equation and satisfies the
stochasticity condition
\begin{align}
\sum_{\bs{i},\bs{j}} \left[S_{I,J}
(\lambda)\right]_{\bs{i},
\bs{j}}^{\bs{i'},\bs{j'}}=1.\label{stoc2}
\end{align}
We can now give the direct proof of \eqref{stoc2} using the explicit
formula \eqref{slnfinal} for the $R$-matrix.

To do that we find it convenient to follow notations of \cite{KMMO16}.
Introduce the function
\begin{align}
\Phi_{q}(\gamma | \beta; \lambda,\mu) &= q^{\xi}
 \left(\frac{\mu}{\lambda}\right)^{|\gamma|}\frac{(\lambda;
q)_{|\gamma|}(\frac{\mu}{\lambda};q)_{|\beta|-|\gamma|}}
{(\mu;q)_{|\beta|}} \prod_{s=1}^{n-1} { \beta_{s} \brack \gamma_{s} }_q,\label{stoc3}\\
\xi &= \sum\limits_{1\leq l<k<n}(\beta_{l}-\gamma_{l})\gamma_{k},\label{stoc3a}
\end{align}
where $\al,\be,\ga,\delta\in \mathbb{Z}_+^{n-1}$, $\lambda,\mu\in\mathbb{C}$.
This function satisfies the following sum rule
\beq
\sum_{\bs{i}} \Phi_{q}(\bs{i} | \bs{j}; \lambda,\mu) = 1.\label{stoc4}
\eeq
Note that the sum in \eqref{stoc4} is always finite since the summand is equal to
zero unless
$\bs{0}\leq \bs{i}\leq \bs{j}$, i.e.
$0\leq i_s\leq j_
s$ for all $s=1,\ldots,n-1$. The relation \eqref{stoc4} can be easily
proved by induction in $n$, see \cite{KMMO16} for details.

Using these definitions and the expansion  \eqref{degen7} 
 ${R}_{I,J}^{(n)}(\lambda)$ can be expressed as
\begin{align}
&\left[{R}_{I,J}^{(n)}(\lambda)\right]_{\bs{i},
\bs{j}}^{\bs{i'},\bs{j'}} =
\delta_{\bs{i}+\bs{j},\bs{i'}+\bs{j'}} \>
q^{(\bs{i'},\bs{j'})-(\bs{i},\bs{j})-
J(|\bs{i}|+|\bs{j}|)+I(|\bs{j}|-|\bs{j'}|)+
\sum_{k>l}(i_{k}j_{l}+j'_{k}i'_{l}-2i_{l}j_{k})}\>\times\nonumber \\
& \sum_{\bs{m}+\bs{n}=\bs{i}+ \bs{j}} \Phi_{q^{2}}
(\bs{j} |\bs{m}; q^{-2J},\lambda^{-2}q^{-I-J})
\Phi_{q^{2}}(\bs{n} | \bs{j'}; \lambda^{2}q^{-I-J},q^{-2J})
q^{2|\bs{n}|J+\sum_{k>l}2(j_{k}n_{l}-j_{l}n_{k})}.\label{stoc5}
\end{align}
where we imply that the sum is taken over $\bs{m},\bs{n}\in\mathbb{Z}_+^{n-1}$ with
the sum $\bs{m}+\bs{n}=\bs{i}+ \bs{j}$ fixed.

Using this presentation of ${R}_{I,J}^{(n)}(\lambda)$ in terms of $\Phi$ 
we can rewrite the expression for matrix elements of $S_{I,J}(\lambda)$ as
\begin{align}
&\left[S_{I,J}(\lambda)\right]_{\bs{i},
\bs{j}}^{\bs{i'},\bs{j'}} =
\delta_{\bs{i}+\bs{j},\bs{i'}+\bs{j'}}
\left(\lambda^{2}q^{I+J}\right)^{|\bs{j}|}\times\nonumber \\
& \sum_{\bs{m}+\bs{n}=\bs{i}+ \bs{j}} \Phi_{q^{2}}(\bs{j}
|\bs{m}; q^{-2J},\lambda^{-2}q^{-I-J})\Phi_{q^{2}}(\bs{n}
| \bs{j'}; \lambda^{2}q^{-I-J},q^{-2J})
q^{-2|\bs{m}|J+\sum_{k>l}2(j_{l}m_{k}-j_{k}m_{l})}\label{stoc6}
\end{align}
This expression can be simplified using symmetries of the function $\Phi$.
Substituting the explicit form of $\Phi$ \eqref{stoc3} one can easily check that
\beq
\Phi_q(\bs{m-j}|\bs{m},\mu/\lambda,\mu)=
\Phi_q(\bs{j}|\bs{m},\lambda,\mu)\>
q^{\>\sum\limits_{k<l}(j_km_l-m_kj_l)}\mu^{-|\bs{j}|}\lambda^{|\bs{m}|}.\label{stoc7}
\eeq
Then we can rewrite \eqref{stoc6} in a  factorized form
\beq
\left[S_{I,J}(\lambda)\right]_{\bs{i},
\bs{j}}^{\bs{i'},\bs{j'}} =
\delta_{\bs{i}+\bs{j},\bs{i'}+\bs{j'}}
\sum_{\bs{m}+\bs{n}=\bs{i}+ \bs{j}}
\Phi_{q^{2}}\left(\bs{m-j}
|\bs{m}; \frac{q^{J-I}}{\lambda^2},\frac{q^{-I-J}}{\lambda^{2}}\right)
\Phi_{q^{2}}\left(\bs{n}
| \bs{j'}; \frac{\lambda^{2}}{q^{I+J}},q^{-2J}\right).\label{stoc8}
\eeq

Now the relation \eqref{stoc2} becomes trivial. Indeed,
for fixed $\bs{i'}, \bs{j'}$ we have
\begin{align}
&\sum_{\bs{i},\bs{j}} \left[S_{I,J}
(\lambda)\right]_{\bs{i},
\bs{j}}^{\bs{i'},\bs{j'}}=
\sum_{\substack{\bs{i}+\bs{j}=\bs{i'}+\bs{j'}\\
\bs{m}+\bs{n}=\bs{i'}+ \bs{j'}}}
\Phi_{q^{2}}\left(\bs{m-j}
|\bs{m}; \frac{q^{J-I}}{\lambda^2},\frac{q^{-I-J}}{\lambda^{2}}\right)
\Phi_{q^{2}}\left(\bs{n}
| \bs{j'}; \frac{\lambda^{2}}{q^{I+J}},q^{-2J}\right)=\nonumber\\
&=\sum_{\bs{m}+\bs{n}=\bs{i'}+ \bs{j'}}
\Phi_{q^{2}}\left(\bs{n}
| \bs{j'}; \frac{\lambda^{2}}{q^{I+J}},q^{-2J}\right)
\sum_{\bs{i}+\bs{j}=\bs{m}+\bs{n}}\Phi_{q^{2}}\left(\bs{m-j}
|\bs{m}; \frac{q^{J-I}}{\lambda^2},\frac{q^{-I-J}}{\lambda^{2}}\right)\nonumber\\
&=\sum_{\bs{m}+\bs{n}=\bs{i'}+ \bs{j'}}
\Phi_{q^{2}}\left(\bs{n}
| \bs{j'}; \frac{\lambda^{2}}{q^{I+J}},q^{-2J}\right)=1,\label{stoc9}
\end{align}
where we  used twice the relation \eqref{stoc4}.

Setting $\lambda=q^{\pm(J-I)/2}$ in \eqref{stoc6} and using relations
\beq
\Phi_q(\bs{i}|\bs{j};1,\mu)=\delta_{\bs{i},0},
\quad \Phi_q(\bs{i}|\bs{j};\mu,\mu)=\delta_{\bs{i},\bs{j}},\label{stoc10}
\eeq
we obtain two nontrivial degenerations of the $R$-matrix $S_{I,J}(\lambda)$
\beq
\left[S^{(1)}(\mu,\nu)\right]_{\bs{i},
\bs{j}}^{\bs{i'},\bs{j'}}\equiv
\left[S_{I,J}
(q^{(J-I)/2})\right]_{\bs{i},
\bs{j}}^{\bs{i'},\bs{j'}}=
\delta_{\bs{i}+\bs{j},\bs{i'}+\bs{j'}}
\Phi_{q^{2}}\left(\bs{i}| \bs{j'}; \mu,\nu\right)\label{stoc11}
\eeq
and
\beq
\left[S^{(2)}(\mu,\nu)\right]_{\bs{i},
\bs{j}}^{\bs{i'},\bs{j'}}\equiv
\left[S_{I,J}
(q^{(I-J)/2})\right]_{\bs{i},
\bs{j}}^{\bs{i'},\bs{j'}}=
\delta_{\bs{i}+\bs{j},\bs{i'}+\bs{j'}}
\Phi_{q^{2}}\left(\bs{j}| \bs{i'}; \nu,\mu\right)\mu^{-|\bs{j}|}\nu^{|\bs{i'}|}
q^{\>2\sum\limits_{k<l}(j_ki'_l-i'_kj_l)},\label{stoc12}
\eeq
where $\mu=q^{-2I},\nu=q^{-2J}$ play the role of (complex) spectral parameters. 
Similar formulas for the $R$-matrix ${R}_{I,J}^{(n)}(\lambda)$ have been already obtained
in \eqref{degen1}-\eqref{degen2}.

We can now derive the formula for the $L$-operator corresponding to the stochastic $R$-matrix
\eqref{stoc1}. First, we choose $I=1$, $J\in\mathbb{Z}_+$ and 
$\bs{i}=\bs{e}_\alpha$,
$\bs{i'}=\bs{e}_\beta$.
Let us notice 
 that the exponent of the $q$-factor in \eqref{stoc1} can be compactly 
written in $n$-component notations as follows
\beq
[\bs{i},\bs{j}]-[\bs{i'},\bs{j'}]
+\sum\limits_{1\leq k<l\leq n}(j_{k}i_{l}-i'_{k}j'_{l})=
{\sum\limits_{k=1}^\al j_k-\sum\limits_{k=\beta}^n j_k'}.\label{stoc13}
\eeq
In particular, for $J=1$  it simplifies to 
\beq
\rho_{\bs{e}_\al,\>\bs{e}_\gamma}^{\bs{e}_\beta,\>\bs{e}_\delta}=
q^{\delta_{\al,\be}\,\epsilon_{\al,\gamma}},\label{stoc14}
\eeq
for $\bs{e}_\al+\bs{e}_\gamma=\bs{e}_\beta+\bs{e}_\delta$ with
\beq
\epsilon_{\al,\gamma}=\begin{cases} 1,&  \al>\gamma,\\
0,& \al=\gamma,\\
-1,& \alpha<\gamma.\end{cases}.\label{stoc15}
\eeq
Let us comment that \eqref{stoc14}
corresponds to the case 
\beq
\rho_{\al,\ga}=q^{\epsilon_{\al,\gamma}}\label{stoc16}
\eeq
in
\eqref{BKMSRmat}. It was shown in \cite{Bazhanov:1990qk} that \eqref{stoc16}
leads to a factorization of the $L$-operators at roots of unity.
It would be interesting to understand further a relation between
stochasticity and factorization of $L$-operators.

We can now derive
the formula for the $L$-operators corresponding to the stochastic $R$-matrix 
$S_{I,J}(\lambda)$. Using \eqref{stoc13} for general $J$ and \eqref{slnlop}
one can write it in terms of Weil generators  \eqref{Weil3} similar to \eqref{slnWeil}
in a compact form
\begin{align}\label{slnWeilS}
L^S_{\al,\be}(\mu) =  
      \mu^{\epsilon_{\al,\be}}\> X^{-1}_\al X_\be\>\left[\mu^{\delta_{\al,\be}} Z_\al\right]
      \prod\limits_{\gamma=1}^{n}Z_\gamma^{\epsilon_{\al,\gamma}}.
\end{align}
It satisfies the algebra
\beq
S_{1,2}(\mu/\nu)L^S_1(\mu)\otimes L^S_2(\nu)=L^S_2(\nu)\otimes L^S_1(\mu)S_{1,2}(\mu/\nu),
\label{stoc17}
\eeq
where $S_{1,2}(\lambda)$ is given by \eqref{stoc1} with $I=J=1$.
This $L$-operator was first obtained in \cite{Bazhanov:1990qk} in a slightly different
form.  The  root of unity condition $q^N=1$ 
used there does not affect 
the local structure of the algebra \eqref{stoc17}.

Choosing the eigenvalue of the operator $\mathcal{Z}$ in \eqref{Weil5}
as  $C$  one can rewrite \eqref{slnWeilS} as
\beq
L_{\al,\be}(\mu) =
      \mu^{\epsilon_{\al,\be}+\delta_{\al,\be}}C\> X^{-1}_\al X_\be\>
      (1-\mu^{-2\delta_{\al,\be}} Z^{-2}_\al)
      \prod\limits_{s=\al+1}^n Z^{-2}_s.\label{slnWeilS1}
      \eeq
This $L$-operator contains two complex parameters $\mu$ and $C=q^J$, 
where $J$ can be identified with the weight of representation. 
As well known
 one can multiply the $L$-operator
\eqref{slnWeilS1}
by  arbitrary  complex parameters $a_i$ (``horizontal'' fields) from the left without affecting
the Yang-Baxter relation. It immediately follows from the property
\beq
[\bs{A}_1\otimes\bs{A}_2,S_{1,2}(\mu)]=0.\label{stoc18},
\eeq
where $\bs{A}=\{a_1,\ldots,a_n\}$.

We can also remove one pair of Weyl operators $Z_1, X_1$ by setting
\beq
Z_1=C\prod_{i=2}^n Z_i^{-1},\quad X_1\equiv 1. \label{stoc19a}
\eeq
Let us introduce
another set of operators
\beq
k_i=q^{-2}Z_{i+1}^{-2},\quad \phi_i^+=X_{i+1}^{-1}(1-Z_{i+1}^{-2}),\quad
\phi_i=X_{i+1}, \quad i=1,\ldots,n-1\label{stoc19}
\eeq
instead of $Z_i,X_i$, $i=2,\ldots,n$. Each set $k_i,\phi_i,\phi^+_i$ forms
 a $q$-oscillator algebra
\beq\label{stock19b}
\phi k=q^2 k \phi,\quad \phi^+ k=q^{-2}k\phi^+,\quad \phi\phi^+-q^2\phi^+\phi=1-q^2.
\eeq

If we now choose
\beq
a_1=-\mu C,\quad a_i=\frac{\mu v}{C}q^{2(i-1-n)},\quad i=2,\ldots,n\label{stoc19c}
\eeq
and make a change of variables
\beq
C=\frac{\sqrt{u v}}{q^n},\quad \mu=q\sqrt{\frac{x}{v}},
\label{stoc20}
\eeq
then we get exactly  the $L$-operator from the recent paper 
by Garbali, De Gier and Wheeler \cite{GGW2016}
\beq
L_{i,j}^{GGW}(x)=a_iL_{i,j}(\mu),
\eeq
with $L_{ij}(\mu)$ given by \eqref{slnWeilS1}. Therefore, the $L$-operator
$L^{GGW}(x)$ corresponds to the standard $U_q(sl(n))$ $L$-operator for symmetric representations
 in the presence of twist and ``horizontal'' fields.

\section{Conclusion}
In this paper we have constructed a new formula of the R-matrix $R(\lambda)$
acting in the tensor product of 
two symmetric representations of the quantum group $U_q(sl_n)$. 
The method is based on calculating the $n$-layer projection 
 of the 3D integrable model introduced in \cite{Bazhanov:2005as,Bazhanov:2008rd,MBS13}.
The final result \eqref{stoc5} can be represented in the  factorized matrix form
with both factors given by a simple product formula \eqref{stoc3}.

The structure of this factorized  representation is quite interesting. The weights of representations
enter the result algebraically with no poles at integer values, 
so the formula equally applies to finite-dimensional and 
infinite-dimensional representations. For integer weights we only need to restrict matrix elements
to basis vectors from finite-dimensional submodules. However, 
the internal sum in \eqref{stoc5} can include vectors beyond finite-dimensional blocks.
 
Following \cite{KMMO16} we also introduced a stochastic $R$-matrix \eqref{stoc1}. 
A factorized representation \eqref{stoc8} makes the proof of stochasticity  almost trivial. 
All matrix elements of the $R$-matrix are positive provided that the condition 
\eqref{convcriteria} is satisfied. Therefore, it defines a discrete time Markov process
with positive probabilities. 

One of the possible directions of future research is to construct stochastic models for 
other Lie algebras. The quantum group approach of \cite{KMMO16} 
suggests that this may be possible and a similar factorization of the $R$-matrix
can exist for other cases.

\section{Acknowledgments}

We would like to thank Jan de Gier, Atsuo Kuniba,  Sergey Sergeev 
for their interest and useful discussions and Vladimir Bazhanov for reading
the manuscript and valuable comments.
\appendixtitleon 
\begin{appendices}
\numberwithin{equation}{section}

\section{}
Here we list standard definitions in $q$-series which we need in the main text
\begin{align}
(a;q)_{\infty} &:= \prod_{i=0}^{\infty} (1-aq^{i}),\\
(a;q)_{n} &:= \frac{(a;q)_{\infty}}{(aq^{n};q)_{\infty}}\label{A1}
\end{align}
\begin{align}
(a_{1},\dots,a_{m};q)_{n} = \prod_{i=1}^{m}(a_{i};q)_{n}\label{A2}
\end{align}
\begin{align}
{n \brack m}_{q} := \frac{(q;q)_{n}}{(q;q)_{n-m}(q;q)_{m}}\label{A3}
\end{align}

We also define a basic hypergeometric series
\begin{align}\label{A4}
{}_{r+1}\phi_{r} \left(\begin{matrix} \left. \begin{matrix} a_{1}, a_{2}, 
\dots,a_{r+1} \\  \phantom{a_1,}b_{1}, \dots,  b_{r}\phantom{w} \end{matrix} \right|  q,x \\
\end{matrix} \right) = \sum_{i\geq0} \frac{(a_1,\ldots,a_{r+1};q)_{i}}{
(q,b_1,\ldots,b_{r};q)_{i}}\> x^{i}.
\end{align}
In the main text we use Heine's transformations of ${}_{2}\phi_{1}$ series
((III.1)-(III-3) in \cite{Gasper})
\begin{align}\label{Heine}
&{}_{2}\phi_{1} \left(\begin{matrix} \left. \begin{matrix} a, b 
\\ c \end{matrix} \right|  q,z\end{matrix} \right)=
\frac{(az,b;q)_\infty}{(c,z;q)_\infty}{}_{2}\phi_{1} \left(\begin{matrix} \left. 
\begin{matrix} c/b,z
\\ az \end{matrix} \right|  q,b\end{matrix} \right)=\nonumber\\
=\frac{(c/b,az;q)_\infty}{(c,z;q)_\infty}&{}_{2}\phi_{1} \left(\begin{matrix} \left.
\begin{matrix} abz/c,b
\\ bz \end{matrix} \right|  q,c/b\end{matrix} \right)=
\frac{(abz/c;q)_\infty}{(z;q)_\infty}{}_{2}\phi_{1} \left(\begin{matrix} \left.
\begin{matrix} c/a,c/b
\\ c \end{matrix} \right|  q,abz/c\end{matrix} \right)
\end{align}
and 
Sears's transformation of terminating ${}_{4}\phi_{3}$ series

\begin{align}\label{Sears}
{}_{4}\phi_{3} \left(\begin{matrix} \left. \begin{matrix} q^{-n},  a,b,c \phantom{I}\\
\phantom{q^{-n},}d, e,  f\phantom{I}\end{matrix} \right|  q,q \\ \end{matrix} \right) &=
\frac{\left(a,\ds\frac{ef}{ab},\frac{ef}{ac};q\right)_{n}}
{\left(e,f,\ds\frac{ef}{abc};q\right)_{n}}\>
{}_{4}\phi_{3} \left(\begin{matrix} \left. \begin{matrix} \ds q^{-n}, & \ds\frac{e}{a},  
&\ds\frac{f}{a},&\ds\frac{ef}{{abc_{\phantom{I}}}}\\
&\ds\frac{ef}{ab},& \ds\frac{ef}{ac}, & \ds\frac{{q^{1-n}}^{\phantom{I}}}{a}\end{matrix} 
\right|  q,q \\ \end{matrix} \right)
\end{align}
provided that $def=abcq^{1-n}$, see (III.16) in \cite{Gasper}.

\end{appendices}

\newcommand\oneletter[1]{#1}
\providecommand{\href}[2]{#2}\begingroup\raggedright\endgroup

\bibliographystyle{utphys}

\begin{thebibliography}{10}

\bibitem{Faddeev:1979}
L.~Faddeev, E.~Sklyanin, and L.~Takhtajan, ``{The Quantum Inverse Problem
  Method. 1},''
{\em Theor.Math.Phys.} {\bfseries 40} (1980) 688--706.

\bibitem{KRS81}
P.~P. Kulish, N.~Y. Reshetikhin, and E.~K. Sklyanin, ``Yang-{B}axter equations
  and representation theory. {I},'' {\em Lett. Math. Phys.} {\bfseries 5}
  no.~5, (1981) 393--403.

\bibitem{KR1981}
P.~P. Kulish and N.~J. Reshetihin, ``Quantum linear problem for the
  sine-{G}ordon equation and higher representations,'' {\em Zap. Nauchn. Sem.
  LOMI,} {\bfseries 101} (1981) 101--110. English translation: J Math Sci
  (1983) {\bf23}:4, 2435-2441.

\bibitem{Fad95}
L.~Faddeev, ``Instructive history of the quantum inverse scattering method,''
  {\em Acta Appl. Math.} {\bfseries 39} no.~1-3, (1995) 69--84. KdV '95
  (Amsterdam, 1995).

\bibitem{Drinfeld86}
V.~G. Drinfeld, ``Quantum groups,'' in {\em Proceedings of the {I}nternational
  {C}ongress of {M}athematicians, {V}ol. 1, 2 ({B}erkeley, {C}alif., 1986)},
  pp.~798--820.
\newblock Amer. Math. Soc., Providence, RI, 1987.

\bibitem{Jim85}
M.~Jimbo, ``A $q$-difference analogue of ${U}({G})$ and the {Y}ang-{B}axter
  equation,'' {\em Lett. Math. Phys.} {\bfseries 10} no.~1, (1985) 63--69.

\bibitem{Bazhanov87}
V.~V. Bazhanov, ``Integrable quantum systems and classical lie algebras.,''
  {\em Comm. Math. Phys.} {\bfseries 113} no.~3, (1987) 471--503.

\bibitem{Jimbo86}
M.~Jimbo, ``Quantum {$R$} matrix for the generalized {T}oda system,'' {\em
  Comm. Math. Phys.} {\bfseries 102} no.~4, (1986) 537--547.

\bibitem{Str79}
Y.~G. Stroganov, ``A new calculation method for partition functions in some
  lattice models,'' {\em Phys. Lett. A} {\bfseries 74} no.~1-2, (1979)
  116--118.

\bibitem{KR87a}
A.~N. Kirillov and N.~Y. Reshetikhin, ``Exact solution of the integrable
  {$XXZ$} {H}eisenberg model with arbitrary spin. {I}. {T}he ground state and
  the excitation spectrum,'' {\em J. Phys. A} {\bfseries 20} no.~6, (1987)
  1565--1585.

\bibitem{Del94}
G.~W. Delius, M.~D. Gould, and Y.-Z. Zhang, ``{On the construction of
  trigonometric solutions of the Yang-Baxter equation},''
  \href{http://dx.doi.org/10.1016/0550-3213(94)90607-6}{{\em Nucl.Phys.}
  {\bfseries B432} (1994) 377--403},
\href{http://arxiv.org/abs/hep-th/9405030}{{\ttfamily arXiv:hep-th/9405030
  [hep-th]}}.

\bibitem{TK92}
V.~N. Tolstoy and S.~M. Khoroshkin, ``Universal ${R}$-matrix for quantized
  nontwisted affine {L}ie algebras,'' {\em Funktsional. Anal. i Prilozhen.}
  {\bfseries 26} no.~1, (1992) 69--71.

\bibitem{BOOS2010}
H.~Boos, F.~G{\"o}hmann, A.~Kl{\"u}mper, K.~S. Nirov, and A.~V. Razumov,
  ``Exercises with the universal {$R$}-matrix,'' {\em J.Phys. A} {\bfseries 43}
  (2010) 415208,
\href{http://arxiv.org/abs/1004.5342}{{\ttfamily arXiv:1004.5342 [math-ph]}}.

\bibitem{Boos11}
H.~Boos, F.~G{\"o}hmann, A.~Kl{\"u}mper, K.~S. Nirov, and A.~V. Razumov, ``On
  the universal {$R$}-matrix for the {I}zergin-{K}orepin model,'' {\em J. Phys.
  A} {\bfseries 44} no.~35, (2011) 355202, 25.

\bibitem{CDS16}
D.~Chicherin, S.~E. Derkachov, and V.~P. Spiridonov, ``{From Principal Series
  to Finite-Dimensional Solutions of the Yang-Baxter Equation},''
  \href{http://dx.doi.org/10.3842/SIGMA.2016.028}{{\em SIGMA} {\bfseries 12}
  (2016) 028},
\href{http://arxiv.org/abs/1411.7595}{{\ttfamily arXiv:1411.7595 [math-ph]}}.

\bibitem{Bazhanov:2005as}
V.~V. Bazhanov and S.~M. Sergeev, ``{Zamolodchikov's tetrahedron equation and
  hidden structure of quantum groups},''
  \href{http://dx.doi.org/10.1088/0305-4470/39/13/009}{{\em J.Phys.} {\bfseries
  A39} (2006) 3295--3310},
\href{http://arxiv.org/abs/hep-th/0509181}{{\ttfamily arXiv:hep-th/0509181
  [hep-th]}}.

\bibitem{Bazhanov:2008rd}
V.~V. Bazhanov, V.~V. Mangazeev, and S.~M. Sergeev, ``{Quantum geometry of
  3-dimensional lattices},''
  \href{http://dx.doi.org/10.1088/1742-5468/2008/07/P07004}{{\em J.Stat.Mech.}
  {\bfseries 0807} (2008) P07004},
\href{http://arxiv.org/abs/0801.0129}{{\ttfamily arXiv:0801.0129 [hep-th]}}.

\bibitem{MBS13}
V.~V. Mangazeev, V.~V. Bazhanov, and S.~M. Sergeev, ``{An integrable 3D lattice
  model with positive Boltzmann weights},''
  \href{http://dx.doi.org/10.1088/1751-8113/46/46/465206}{{\em J.Phys.}
  {\bfseries A46} (2013) 465206},
\href{http://arxiv.org/abs/1308.4773}{{\ttfamily arXiv:1308.4773 [math-ph]}}.

\bibitem{Man14}
V.~V. Mangazeev, ``{On the Yang-Baxter equation for the six-vertex model},''
  {\em Nuclear Phys. B} {\bfseries 882} no.~1, (2014) 70--96,
  \href{http://arxiv.org/abs/1401.6494}{{\ttfamily arXiv:1401.6494 [math-ph]}}.

\bibitem{BCG16}
A.~Borodin, I.~Corwin, and V.~Gorin, ``{Stochastic six-vertex model},''
  \href{http://dx.doi.org/10.1215/00127094-3166843}{{\em Duke Math. J.}
  {\bfseries 165} no.~3, (2016) 563--624}.

\bibitem{Bor14}
A.~Borodin, ``{On a family of symmetric rational functions},''
\href{http://arxiv.org/abs/1410.0976}{{\ttfamily arXiv:1410.0976 [math.CO]}}.

\bibitem{CP16}
I.~Corwin and L.~Petrov, ``{Stochastic higher spin vertex models on the
  line},'' \href{http://dx.doi.org/10.1215/00127094-3166843}{{\em Comm. Math.
  Phys.} {\bfseries 343} no.~2, (2016) 651--700}.

\bibitem{BP16}
A.~Borodin and L.~Petrov, ``{Higher spin six vertex model and symmetric
  rational functions},''
\href{http://arxiv.org/abs/1601.05770}{{\ttfamily arXiv:1601.05770 [math.PR]}}.

\bibitem{Povol13}
A.~M. Povolotsky, ``On the integrability of zero-range chipping models with
  factorized steady states,'' {\em J. Phys. A} {\bfseries 46} no.~46, (2013)
  465205, 25.

\bibitem{Man14b}
V.~V. Mangazeev, ``{$Q$-operators in the six-vertex model},''
  \href{http://dx.doi.org/10.1016/j.nuclphysb.2014.06.024}{{\em Nucl. Phys.}
  {\bfseries B886} (2014) 166--184},
\href{http://arxiv.org/abs/1406.0662}{{\ttfamily arXiv:1406.0662 [math-ph]}}.

\bibitem{KMMO16}
A.~Kuniba, V.~V. Mangazeev, S.~Maruyama, and M.~Okado, ``{Stochastic $R$ matrix
  for $U_q(A_n^{(1)})$},''
\href{http://arxiv.org/abs/1604.08304}{{\ttfamily arXiv:1604.08304 [math.QA]}}.

\bibitem{GGW2016}
A.~Garbali, J.~{De Gier}, and M.~Wheeler, ``{A new generalisation of Macdonald
  polynomials},'' \href{http://arxiv.org/abs/1605.07200}{{\ttfamily
  arXiv:1605.07200 [math-ph]}}.

\bibitem{DerMan2010}
S.~E. Derkachov and A.~N. Manashov, ``A general solution of the {Y}ang-{B}axter
  equation with the symmetry group {${\rm SL}(n,\Bbb C)$},'' {\em Algebra i
  Analiz} {\bfseries 21} no.~4, (2009) 1--94.

\bibitem{Cherednik1980}
I.~V. {{C}}herednik, ``On a method of constructing factorized {$S$}-matrices in
  terms of elementary functions,'' {\em Teoret. Mat. Fiz.} {\bfseries 43}
  no.~1, (1980) 117--119.

\bibitem{Kulish:1980ii}
P.~P. Kulish and E.~K. Sklyanin, ``Solutions of the {Y}ang-{B}axter equation,''
{\em J. Sov. Math.} {\bfseries 19} (1982) 1596--1620.

\bibitem{Perk:1981nb}
J.~H.~H. Perk and C.~L. Schultz, ``New families of commuting transfer matrices
  in q-state vertex models,''
{\em Phys. Lett.} {\bfseries A84} (1981) 407--410.

\bibitem{Bazhanov:1990qk}
V.~V. Bazhanov, R.~M. Kashaev, V.~V. Mangazeev, and Y.~G. Stroganov,
  ``{$(Z_N\times)^{n-1}$ generalization of the chiral Potts model},''
\href{http://dx.doi.org/10.1007/BF02099497}{{\em Commun. Math. Phys.}
  {\bfseries 138} (1991) 393--408}.

\bibitem{Gasper}
G.~Gasper and M.~Rahman, \href{http://dx.doi.org/10.1017/CBO9780511526251}{{\em
  Basic hypergeometric series}}, vol.~96 of {\em Encyclopedia of Mathematics
  and its Applications}.
\newblock Cambridge University Press, Cambridge, second~ed., 2004.

\end{thebibliography}
\end{document}